\documentclass[twocolumn,prb,aps,superscriptaddress,showpacs,floatfix]{revtex4}

\usepackage{amssymb}
\usepackage{dcolumn}
\usepackage{epsfig}
\usepackage{graphicx}
\usepackage{epstopdf}
\usepackage{float}
\usepackage[tbtags]{amsmath}
\newlength{\figwidth}
\newlength{\figwidthb}
\begin{document}

\title{Magnetic states at the surface of $\alpha-$\rm Fe$_{2}$O$_{3}$ thin films doped with Ti, Zn or Sn}
\author{David~S. Ellis}
\affiliation{Department of Materials Science and Engineering, Technion-Israel Institute of Technology, Haifa 32000, Israel}
\author{Eugen Weschke}
\affiliation{Helmholtz-Zentrum Berlin für Materialen und Energie, Albert-Einstein-Strasse 15, 12489 Berlin, Germany}
\author{Asaf Kay}
\affiliation{Department of Materials Science and Engineering, Technion-Israel Institute of Technology, Haifa 32000, Israel}
\author{Daniel~A. Grave}
\affiliation{Department of Materials Science and Engineering, Technion-Israel Institute of Technology, Haifa 32000, Israel}
\author{Kirtiman~Deo Malviya}
\affiliation{Department of Materials Science and Engineering, Technion-Israel Institute of Technology, Haifa 32000, Israel}
\author{Hadar Mor}
\affiliation{Department of Materials Science and Engineering, Technion-Israel Institute of Technology, Haifa 32000, Israel}
\author{Frank M. F. de Groot}
\affiliation{Department of Inorganic Chemistry and Catalysis, Debye Institute, Utrecht University, Sorbonnelaan 16, 3584 CA Utrecht, The Netherlands}
\author{Hen Dotan}
\affiliation{Department of Materials Science and Engineering, Technion-Israel Institute of Technology, Haifa 32000, Israel}
\author{Avner Rothschild}
\affiliation{Department of Materials Science and Engineering, Technion-Israel Institute of Technology, Haifa 32000, Israel}

\email{avner@mt.technion.ac.il}

\date{\today }

\begin{abstract}
The spin states at the surface of epitaxial thin films of hematite, both undoped and doped with 1\% Ti, Sn or Zn, respectively, were probed with x-ray magnetic linear dichroism (XMLD) spectroscopy.  Morin transitions were observed for the undoped ($T_M\approx200$ K) and Sn-doped ($T_M\approx$300 K) cases, while Zn and Ti-doped samples were always in the high and low temperature phases, respectively.  In contrast to what has been reported for bulk hematite doped with the tetravalent ions Sn$^{4+}$ and Ti$^{4+}$, for which $T_M$ dramatically decreases, these dopants substantially increase $T_M$ in thin films, far exceeding the bulk values.   The normalized Fe $L_{II}$-edge dichroism for $T<T_M$ does not strongly depend on doping or temperature, except for an apparent increase of the peak amplitudes for $T<100$ K.  We observed magnetic field-induced inversions of the dichroism peaks.  By applying a magnetic field of 6.5 T on the Ti-doped sample, a transition into the $T>T_M$ state was achieved.  The temperature dependence of the critical field for the Sn-doped sample was characterized in detail.  It was demonstrated the sample-to-sample variations of the Fe $L_{III}$-edge spectra were, for the most part, determined solely by the spin orientation state.  Calculations of the polarization-depedent spectra based on a spin-multiplet model were in reasonable agreement with the experiment and showed a mixed excitation character of the peak structures.

\end{abstract}

\maketitle

\label{sect:intro} 
Hematite ($\alpha$-$\rm Fe{_2}O_{3})$ is considered to be a promising material for photoelectrochemical (PEC) cells for solar water splitting~\cite{Sivula11,Kay06,Tamirat16}, whereby solar energy can drive hydrogen-production reactions in water~\cite{Fujishima72}.  Introducing impurities, or doping, as a source of free charge carriers is a crucial ingredient to making feasible devices~\cite{Malviya16,Yatom16}.  The models and tools of semiconductor device physics are being increasingly used in the ongoing effort to develop such cells using hematite photo-anodes \cite{Kay16}.  Yet, the electron-electron interactions arising from its $\rm Fe^{3+}$ 3$d$-orbitals, mediated by oxygen atoms, make hematite a strongly correlated system, as evidenced by its antiferromagnetism (AF).  In addition to the effect of doping on transport properties in the usual way, the doping can also be used to manipulate the magnetic structure. As to which extent this is also important for the device performance has yet to be explored.\

\begin{figure}[htbp]
	\centering
	\epsfig{file=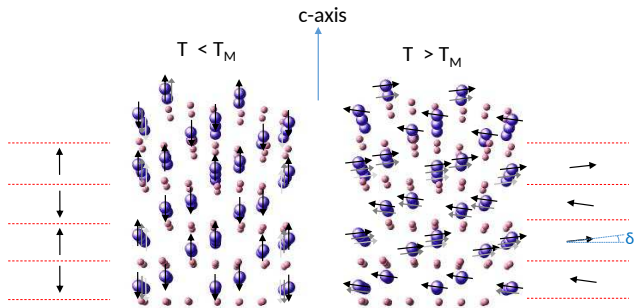,height=1.7in,keepaspectratio}
	\vspace*{-5mm}
	\caption{(Color Online).  The spin ordering in hematite for the magnetic state for $T<T_M$ (left) and $T>T_M$ (right).  The iron atoms are represented by the large blue spheres, and arrows pointing in their spin directions, and the small red spheres are oxygen atoms separating the Fe planes.  Within each plane, the spin directions are the same, as indicated by the stacks of arrows on each side. For $T>T_M$, the spins cant out of the basal plane with an angle $\delta$, as shown.   Made with the help of freely available \textit{Balls and Sticks} software \cite{Ozawa04}.  }
	\label{fig:spins}
\end{figure}

A magnetic transition, known as the Morin transition, takes place upon raising the temperature past $T_M$ (=265 K for bulk, undoped hematite), which changes the spin-ordering from pure AF to a superposition of AF with weak ferromagnetism \cite{Morin50}.  Polarized neutron diffraction \cite{Shull51}, magnetic resonance \cite{Anderson54,Kumagi55} and magnetostriction measurements \cite{Urquhart56} have since confirmed this and revealed that below $T_M$ the spins are aligned antiferromagnetically along the $c$-axis, with stacked $c$-planes having alternating up and down spins. The spin arrangements are illustrated in fig.~\ref{fig:spins}.  Above $T_M$, the spins are rotated $\sim$90$^{\circ}$ to lie in the $c$-plane, with AF order still maintained along the $c$ direction, but now with respect to the in-plane spin. In this state, referred to here as the ``Morin phase", the spins tilt slightly out of plane in the same out-of-plane direction for all of the stacked basal planes, which is the origin of the weak ferromagnetism. The state below $T_M$ we refer to as the ``AF phase" for brevity.  The magnetism was seen to significantly affect other physical properties, such as electrical resistance\cite{Nakau60,Meng13}, and Hall effect\cite{Morin51}, which is anomolous in the Morin state.  Recently the Hall effect was observed to behave much more normally in the AF state ($T<T_M$), in lightly Si-doped bulk single crystals oriented with (0001) surfaces \cite{Rettie16}.\

Doping is known to alter the transition temperature, or even suppress the Morin transition altogether.  Several studies of doped bulk samples have been performed\cite{Morrish94}, although magnetic measurements of oriented thin films has been relatively recent.  The tetravalent Ti and Sn dopants have been observed to drastically decrease $T_M$ in bulk hematite for concentrations as little as 0.3\% \cite{Curry65,Besser67,Kotyuzhanskii79,Ericsson86}, consistent with the observed anomalous Hall effect\cite{Morin51}.  The $T_M$ decrease for the tetravalent dopants was attributed to increased single ion anisotropy\cite{Artman65}, mainly of charge-compensating Fe$^{2+}$ ions\cite{Flanders65}. More recently, Zhao et al. \cite{Zhao11}, measured Ti-doped hematite thin films, grown on (0001)-oriented sapphire substrates, which exhibited a much less anomalous (yet not completely consistent) Hall response.  Taking into account Ref.~\cite{Rettie16}, the latter result may hint that the thin films were in the AF phase, rather than the Morin phase, suggesting an entirely opposite effect of the Ti dopant on the magnetism in thin films, as compared to the bulk.

Here we measure the magnetic state of undoped and doped, thin-film hematite samples.  The dopants studied were chosen so as to include two tetravalent ions, Sn$^{4+}$, Ti$^{4+}$, a divalent Zn$^{2+}$, and an undoped film.  A doping level of 1\% is of the order used for many of the previous studies described above.  In addition to the above interesting behavior, these are typical donors and acceptors, and doping levels used in designs of hematite photoanode devices\cite{Kay16,Malviya16}.  We initially attempted SQUID magnetometry, but as a bulk probe it was susceptible to contamination from the much more voluminous substrate or holder, which we found to be much greater than the total magnetization of our thin film. A similar difficulty was encountered in the study of magnetism in Mn-doped ZnO layers on sapphire substrate \cite{Mofor07}, and the problem worsens upon going to even thinner layers.  It should be noted that recently Shimomura et. al.\cite{Simomura15} successfully observed the Morin transition in thin films using a SQUID, but even with pure substrate and sensitive instrument, a background needed to be subtracted.  We therefore used x-ray magnetic linear dichroism (XMLD) \cite{Thole85}, a method particularly suited for thin films or surfaces of hematite\cite{Kuiper93}.  XMLD allows one to ignore the backround and probe the iron atoms exclusively.   \

\begin{figure}[hbp]
	\centering
	\vspace*{-3mm}
	\epsfig{file=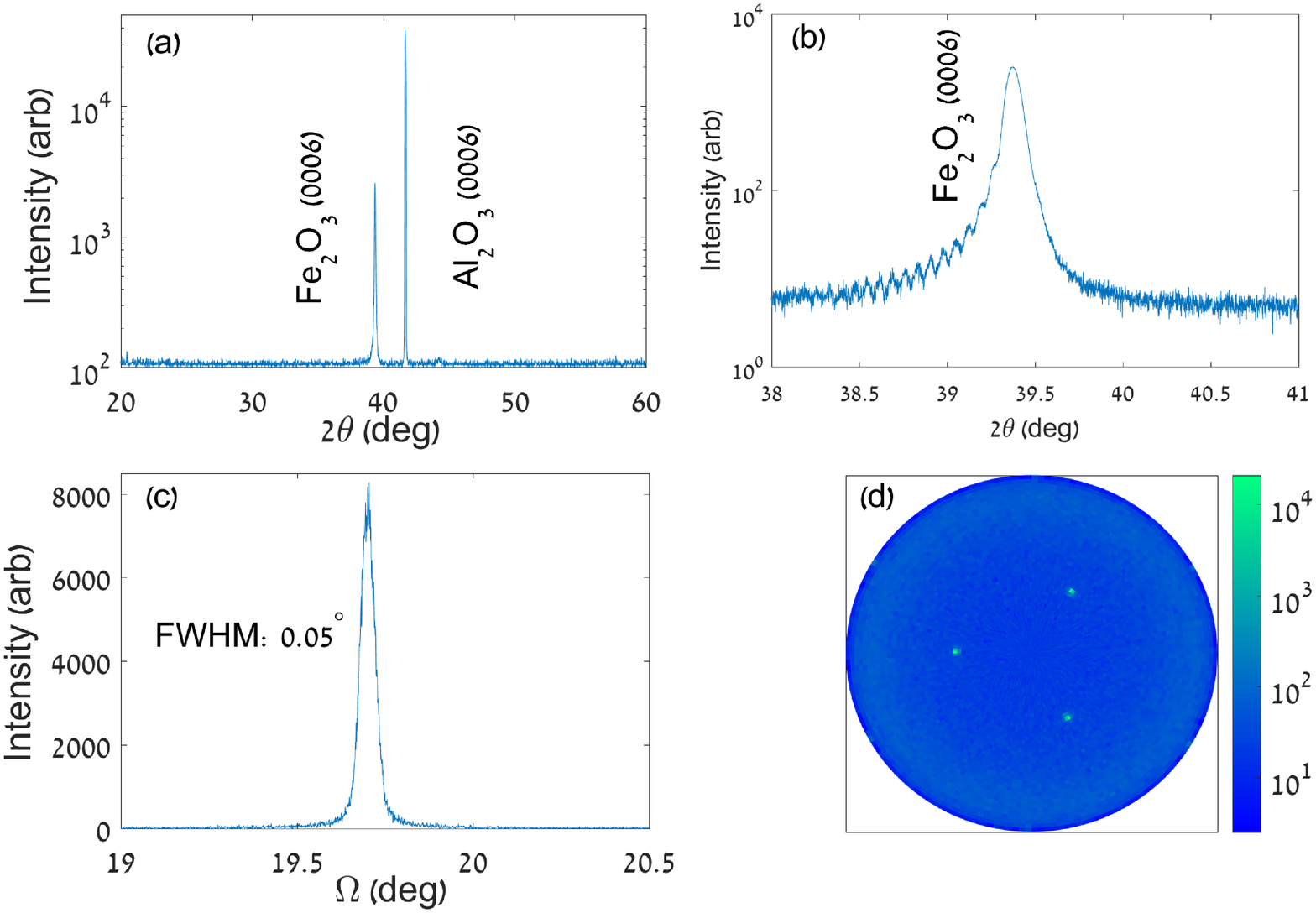,height=2.5in,keepaspectratio}
	\vspace*{-3mm}
	\caption{(a) $\theta$-2$\theta$ x-ray diffraction pattern showing $c$-axis orientation of hematite thin film grown on $c$-plane sapphire. (b) high resolution x-ray diffraction pattern of hematite (0006) reflection revealing the existence of Laue oscillations (c) Rocking curve of (006) peak having a width of 0.05$^{\circ}$. (d) Pole figure of the (1 0 4) reflection showing heteroepitaxial in-plane alignment of hematite films on sapphire substrate with small in-plane mosaic spread.}
	\label{fig:xrd}
\end{figure}

By using a controlled epitaxial process consistent for all of the samples, the effect of doping could be better observed with minimal variation of other factors, such as surface morphology or contamination with random impurities.  The hematite films were prepared according to a previous report \cite{Grave16}.  The films were deposited from individually doped targets, by pulsed laser deposition on $c$-plane (0001) sapphire substrates, at a set point temperature of 800 $^{\circ}$C under an oxygen partial pressure of 10 mTorr, using 20,000 pulses for a thickness of $\sim$150 nm. Fig.~\ref{fig:xrd}(a)-(d) summarizes the characterization of an undoped sample fabricated with this process, using a Rigaku Smartlab x-ray diffractometer.  The existence of Laue oscillations in Fig.~\ref{fig:xrd}(b) indicates excellent crystallinity and smooth, well ordered surfaces, and the narrow rocking curve width of 0.05$^{\circ}$ in Fig. \ref{fig:xrd}(c) shows the excellent crystal mosaic, and single domain is indicated by the pole graph in Fig. \ref{fig:xrd}(d).  The doped films were of similar crystal and surface quality.  The Ti- and Sn-doped films were determined to be strong n-type, the undoped film weak n-type, and the Zn-doped film weak p-type~\cite{Kay16}.  The lattice constants, determined from least square fitting of the scattering angles of six diffraction peaks, are tabulated in Table 1.  However, for a complete description of the structure relevant for the magnetic interactions, the displacement \textit{W} of the iron atoms out of the basal plane would also be needed \cite{Artman65}, which unfortunately we could not measure using our x-ray diffractometer.\  

\begin{table}[h]
	\centering
	\begin{tabular}{|l|l|r|l|}
		\hline
		Sample & $a (\AA)$ & $c (\AA)$ \\
		\hline
		undoped 150 nm & 5.045 $\pm$ 0.002& 13.730 $\pm$ 0.002\\
		\hline
		1\% Ti-doped 150 nm & 4.992 $\pm$ 0.002 & 13.859 $\pm$ 0.003\\
		\hline
		1\% Sn-doped 150 nm & 5.039 $\pm$ 0.005   & 13.754 $\pm$ 0.006 \\
		\hline
		1\% Zn-doped 150 nm & 5.002 $\pm$ 0.003   & 13.850 $\pm$ 0.004  \\
	    \hline
	   bulk hematite (Ref.~\cite{Steinwehr67}) & 5.03490(9) & 13.7524(18)  \\
	    \hline
	\end{tabular}
	\caption{\label{tab:lattice } The hexagonal lattice constants of the four samples.  For comparison, the last entry in the table is for bulk hematite as reported in Ref.~\cite{Steinwehr67}}
\end{table}

\begin{figure}[hbp]
	\centering
	\epsfig{file=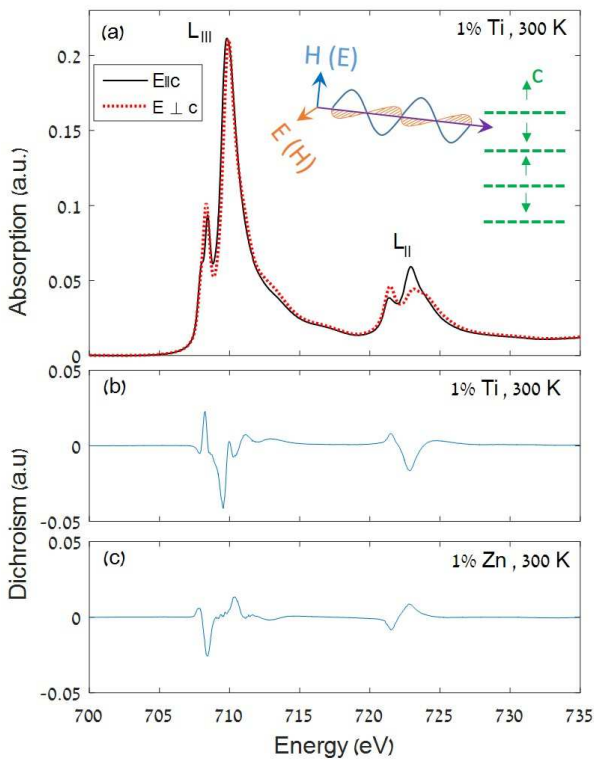,height=4.0in,keepaspectratio}
	\vspace*{-2mm}
	\caption{(Color Online) (a) The Fe $L_{II}$ and $L_{III}$ x-ray absorption spectra for the electric field of the incident x-ray approximately parallel and perpendicular to the $c$-axis of the 1\% Ti-doped sample at $T$=300 K.  The linear polarizations of the near-grazing incident x-ray with respect to the $c$-axis of the sample is indicated. (b) The linear dichroism spectrum, defined as the difference between the two curves in (a).  (c) The room temperature dichroism of the Zn-doped sample.}
	\label{fig:absexam}
\end{figure}

\begin{figure*}[tbp]
	\centering
	\epsfig{file=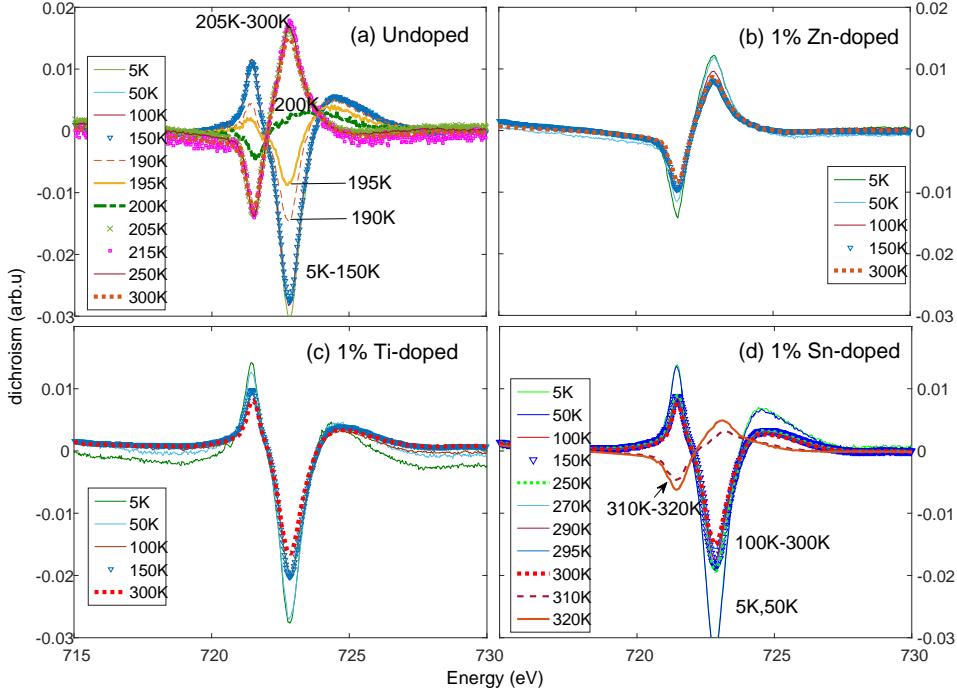,height=3.8in,keepaspectratio}
    \vspace*{-3mm}
	\caption{(Color Online)  The dichroism spectra at the Fe $L_{II}$-edge, measured at the indicated temperatures for the (a) undoped, (b) 1\% Zn-doped, (c) 1\% Ti-doped, and (d) 1\% Sn-doped samples. }
	\label{fig:Tdep}
\end{figure*}

\begin{figure}[htbp]
	\centering
	\epsfig{file=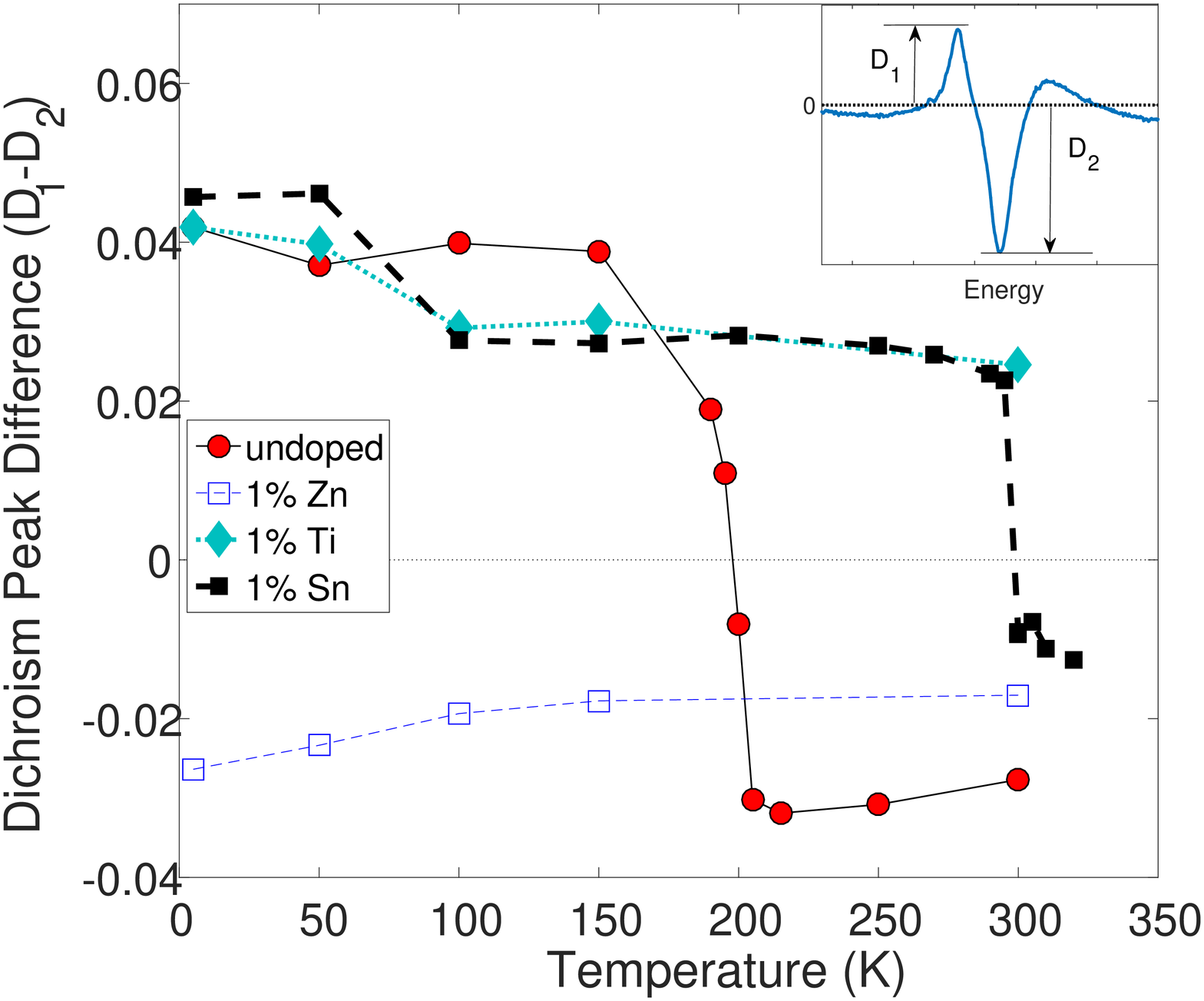,height=2.7in,keepaspectratio}
	\vspace*{-5mm}
	\caption{(Color Online) The difference of peak height values between the first and second main peaks of the Fe $L_{II}$-edge dicrhoism spectra, determined relative to zero as shown in the inset (below zero is considered as a negative peak height). They are plotted as a function of temperature for the four samples.}
	\label{fig:peakheights}
\end{figure}

The magnetic ordering of the samples were determined from their XMLD spectra, which compares the x-ray absorption (XAS) spectra for polarizations of the incoming x-ray beam which are parallel and perpindicular to the ordered moments \cite{Thole85}.  For hematite, the ratios of the two peaks of the Fe $L_{II}$ edge  exhibit a clear change between the AF and Morin states \cite{Kuiper93,Gota01,Park13}.  Our measurements were done with a high-field diffractometer at the UE46-PMG1 beamline at the BESSY II synchrotron, using the total electron yield (TEY) method. This method is highly surface sensitive, since detected electrons originate no deeper than a few nm from the surface.  Therefore, our results below implicitly refer to the surface properties.  The x-ray beam was incident at an angle of 20$^{\circ}$ to the sample surface, such that horizontally and vertically polarized incident x-rays were approximately parallel and perpendicular to the $c$-axis, respectively.  An initial check verified that rotating the sample about the $c$-axis did not affect the normalized dichroism, consistent with a three-fold in-plane magnetic domain structure.  Nevertheless, all of the samples were mounted with consistent in-plane orientation.   Consistency of repeated measurements at each polarization was checked to ensure that there were no surface charging effects in the TEY signal.  Nominally, a small magnetic field of 0.05 T was applied to the sample (herein we refer to the field in units of magnetic induction in air, so that 1 T corresponds to 10 kG), which was increased for field-dependence measurements.\
  
Fig.~\ref{fig:absexam}(a) shows a typical normalized XAS spectra at the Fe $L$-edge for the Ti-doped hematite film, for two polarizations, and their difference plotted in Fig.~\ref{fig:absexam}(b), which is the XMLD spectrum. The XAS spectra were each normalized by first subtracting the value at 700 eV, and then dividing by the resultant area between 700 and 735 eV.  All of the spectra, for all samples and all temperatures/ fields were normalized in this manner, in order to allow for a consistent comparison of the dichroism.  As an example of how the dichroism can change, we plot it for the Zn-doped sample in Fig.~\ref{fig:absexam}(c).  Comparison of panels (b) and (c) show that the two prominent XMLD peaks of the $L_{II}$-edge, between 720-725 eV, invert sign.  In contrast, while the $L_{III}$-edge XMLD features, in the 705-715 eV range, exhibit some inversion, the structure is generally more intricate.  A rigorous analysis of the $L_{III}$-edge, as well as the O $K$-edges, can provide additional information about valence and local environment \cite{Miedema13,Braun12,Braun16}.   However, as will be shown, the variation amongst our samples of the features in the $L_{III}$-edge, appear to depend primarily on the spin-ordering state, which can be determined by the  $L_{II}$-edge.   Here we focus on the $L_{II}$-edge which provides information on the spin directions.\

The temperature dependence of the dichroism spectra at the Fe $L_{II}$-edge is plotted for each sample, in Fig.~\ref{fig:Tdep} (a)-(d).  The magnitudes of the dichroism peaks, especially below $T_M$, are for the most part sample independent.  This proves that the normalization procedure removes the effect of conductivity on the TEY spectra, which varies enormously between undoped and Ti-doped samples.   In the undoped and Sn-doped samples, an inversion of the signs of the dichroism of the first peaks occurs at $T$=200 K and $T$=300 K respectively.  The transition temperature for the undoped film surface is well below that of the bulk $T_M$=265 K, but it is known that $T_M$ varies with thickness in a non-trivial way, depending on strain and/or atomic site vacancies in the lattice \cite{Gota01,Park13,Aragon16}, and our value interpolates well with the recent study of Shimomura et al. for similar thicknesses \cite{Simomura15}.  Inspection of Fig. \ref{fig:Tdep}(a) and (d) shows that the transition itself is evident by a flipping of the signs of the peaks, which is relatively sharp in temperature in comparison to the changes away from the transition.\

In order to more quantitatively characterize the dichroism, we plot the difference between the first and second dichroism peak heights (with respect to zero), as a function of temperature in Fig.~\ref{fig:peakheights}.  The Ti and Zn-doped samples do not exhibit a transition, with the XMLD of the Ti-doped sample always in the low-temperature side, and that of the Zn-doped sample always on the high temperature side of the transition.  This suggests that the Zn-doped sample is always in the Morin phase, which would be consistent with a recent M\"ossbauer study on bulk sample \cite{Varshney14}.  That Ti doping, in contrast, favors the AF phase is consistent with the (relatively) normal Hall voltage dependence on magnetic field in Ti-doped thin films \cite{Zhao11}, taking into account the comparative study of Hall effect above and below $T_M$ \cite{Rettie16}.  Comparing samples of the different dopants, there was no obvious trend of $T_M$ with lattice constant, as the Fe displacement structure parameter $W$ is likely to play a key role\cite{Artman65}.  A synchrotron study such as EXAFS or high-resolution diffraction on a dedicated beamline, would be needed to determine $W$ for the thin films.   


In Fig.~\ref{fig:peakheights}, the Fe $L_{II}$-edge dichroism of the Ti- and Sn-doped samples are overlapping over a wide temperature range for $T<T_M$, even close to the Morin transition of the Sn-doped sample.  The Ti-doped sample is rather far from a transition in this temperature range, as will be estimated from the field dependence below.  Both of the Sn- and Ti-doped curves seem to have an abrupt change in slope for temperatures below 100 K, whereas the peak changes relatively slowly above $T$=100 K, which is also clear from Fig.~\ref{fig:Tdep} (c) and (d).  Such a kink does not appear clearly in the undoped sample, but is also evident for the Zn-doped sample.  Its origin is uncertain, and while we cannot yet rule out some unknown artifact, it may hint at a possible new phase at low temperature, and requires further investigation.  In contrast to the AF phase, the magnitudes of the peaks in the Morin phase appear to have greater variation with dopant, more than doubling for the undoped sample (Fig.~\ref{fig:Tdep}(a)) as compared to 1\% Sn-doped (Fig.~\ref{fig:Tdep}(d)).  Since the spins in the AF phase are pointing exactly along the c-axis, but in the Morin phase there is an additional degree of freedom of canting angle, we posit that the observed variation in the dichroism could be a consequence of the canting angle variation.  \\

\begin{figure}[htbp]
	\centering
	\epsfig{file=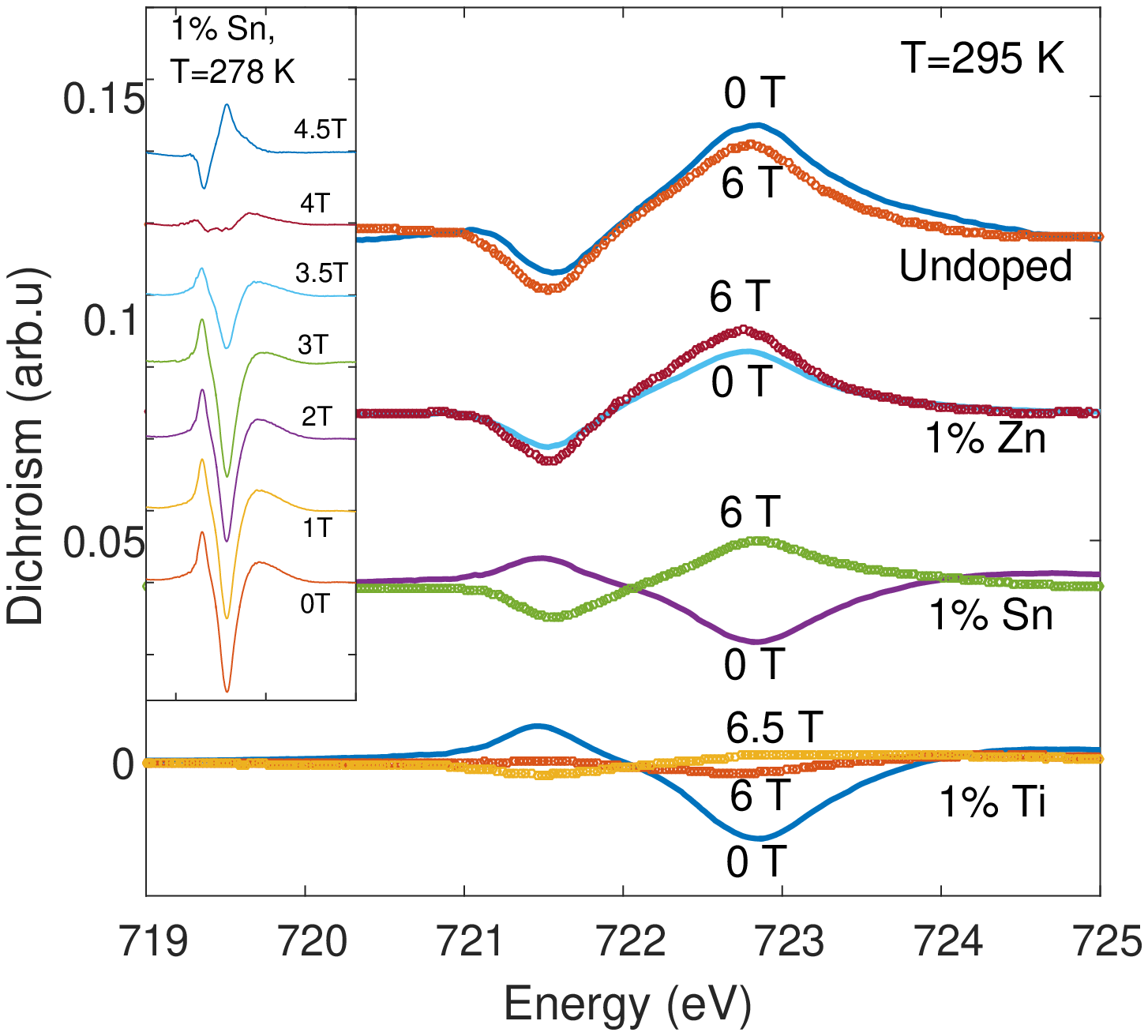,height=3.3in,keepaspectratio}
	\vspace*{-5mm}
	\caption{(Color Online) Effect of applying high magnetic fields (20$^{\circ}$ to the $c$-axis) on the dichroism spectra of each of the samples, at $T$=295 K.  For clarity the spectra of each sample are displaced along the vertical axis.  The fields are as indicated, with solid curves corresponding to B=0.05 T (approximately zero), and the circles corresponding to the high magnetic fields.  The inset shows a typical example of the progression of the spectra with increasing field in the 1\% Sn-doped sample, with a magnetic phase transition induced at B=4 T, for $T$=278 K.}
	\label{fig:Hdep}
\end{figure}

A strong magnetic field parallel to the spins in the AF phase is known to induce a spin-flop transition in hematite \cite{Besser64,Foner65}, whereby the spins rotate perpendicular to the field (corresponding to the Morin phase).  However, to our knowledge, this was never measured with XMLD.  It would be expected that a strong enough field should flip of the signs of the measured dichroism of samples initially in the pure AF phase.  The magnetic field dependence ($H$) of the dichroism at $T$=295 K was first checked by comparing the spectra for $H$=6 T applied 20$^{\circ}$ to the normal to the surface, and for nearly zero nominal field of $H$=0.05 T.  As shown in fig.~\ref{fig:Hdep}, the magnetic field was indeed able to induce an inversion of the dichroism peaks for the Sn-doped sample, and also flattened the Ti-doped spectra.  Increasing $H$ further, to 6.5 T, does begin to invert the sign of the dichroism of the Ti sample.  The undoped and Zn-doped samples were not affected magnetic field, since they are already in the Morin state at $T$=295 K.\

\begin{figure}[htbp]
	\centering
	\epsfig{file=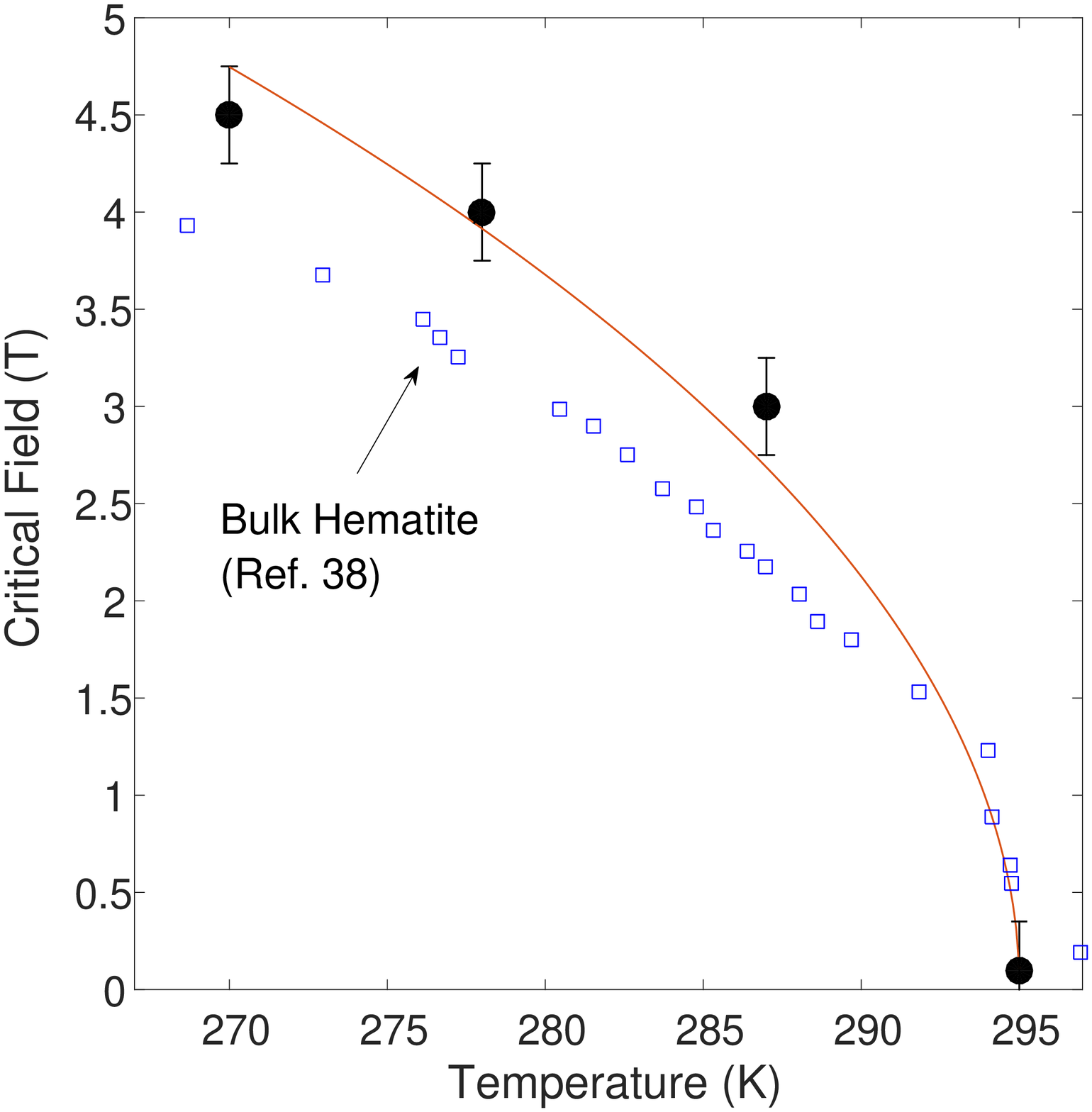,height=3.7in,keepaspectratio}
	\vspace*{-5mm}
	\caption{(Color Online) Critical field as a function of temperature for the 1\% Sn-doped sample (large circles), fit to square-root function (solid line).  The result of Foner and Shapira for bulk hematite \cite{Foner69}, shifted along the temperature axis by 35 K, which is the difference of $T_M$'s between the two samples  (small squares).}
	\label{fig:Hc}
\end{figure}

The critical field required to induce transitions in the Sn-doped sample was measured at each temperature, by incrementing the magnetic field until a transition was found.  An example of the progression of the spectra is plotted in the inset of Fig.~\ref{fig:Hdep}, which indicates a width of the transition with field of approximately $\pm$0.5 T.  The critical field $H_c(T)$, determined from the field of flat dichroism at the $L_{II}$-edge, is plotted in fig.~\ref{fig:Hc}.  The data fits well to a $H_c(T)=a \cdot (T_M-T)^{\frac{1}{2}}$ vs. $T$ curve.  It is interesting that the $H_c(T)$ curve is of similar magnitude to that measured of bulk hematite by Foner and Shapira \cite{Foner69}, also plotted in fig.~4.  Similar $H_c(T)$ curves were also observed among doped bulk samples by Besser et al \cite{Besser67}, who further used the observed low-temperature critical fields to extract the anisotropy fields and accurately predict their respective $T_M$'s \cite{Artman65}. Although we did not measure $H_c(T)$ for our other samples, this similarity suggests that they should likewise be expected to share a similar curve relative to their respective $T_M$'s.   A recent magnetization study of hematite thin films by Pati et al.\cite{Pati17} showed generally the same magnitude of slope as in Fig.~\ref{fig:Hdep}. Using the $H_c(T)$ curve of Fig.~\ref{fig:Hc} as a guide, one can predict $T_M$ to be around 320-330 K for the Ti-doped sample, based on $H_c(295K)$=6 T from Fig.~\ref{fig:Hdep}.\\


  \begin{figure*}[tbp]
  	\centering
  	\epsfig{file=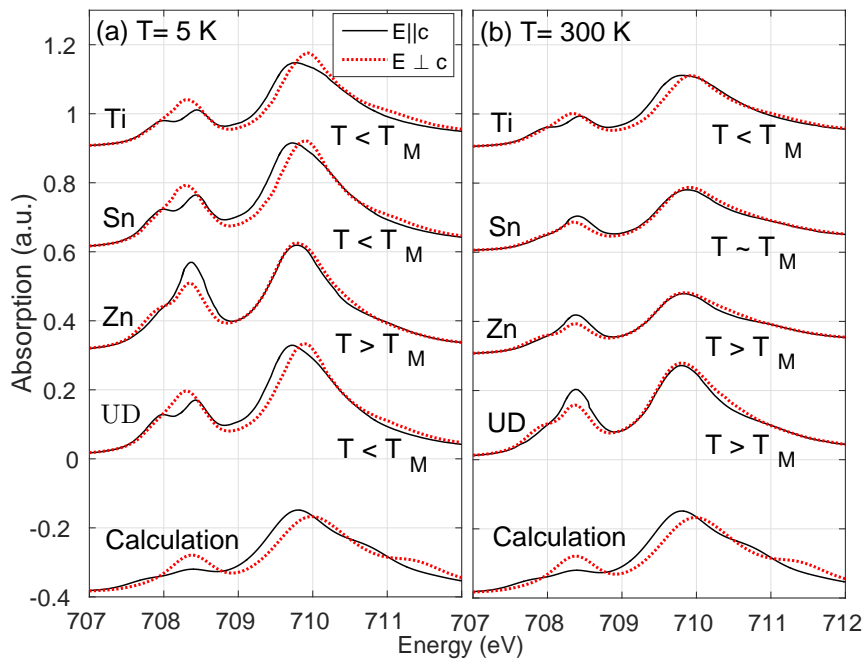,height=3.4in,keepaspectratio}
  	\vspace*{0mm}
  	\caption{(Color Online) Absorption spectra at the Fe $L_{III}$-edge, for polarizations approximately parallel and perpindicular to the crystallographic $c$-axis, for the four samples with dopants as indicated, for (a) $T$=5 K, and (b) $T$=300 K.  The spectra for each sample are offset along the vertical axis for clarity.  The spin state, as determined from the analysis of the $L_{II}$-edge, is indicated by the labels of temperature relative to the Morin transition.   The bottom curves (identical for (a) and (b)) are from calculations using the program CTM4XAS as described in the text.}
  	\label{fig:L3}
  \end{figure*}

Lastly, it is worthwhile to re-visit the $L_{III}$-edge absorption spectra.    The two panels (a) and (b) for Fig.~\ref{fig:L3} show the polarized spectra for $T$=5 K and $T$=300 K, along with the magnetic state indicated for each as determined by the above $L_{II}$-edge analysis, and with calculations that will be described below.   Inspection of the experimental curves in Fig.~\ref{fig:L3} reveals that, irrespective of particular temperature or dopant, the qualitative properties of each spectral pair are almost completely determined by the magentic state.  The $T<T_M$ spectra all exhibit a double peak at ~708-708.5 eV for the $c$-polarized absorption (black lines), and an apparently single, broad peak in the same range for the basal-plane polarized absorption (red lines), which appears to encompass the former peaks for all samples/temperatures in this state.  The 709.5-710 eV region in this state is characterized by a peak for $E\perp$$c$ which is shifted 0.2 eV with respect to the peak for $E\parallel$$c$.  In contrast, the peaks of the $T>T_M$ spectra in this range coincide for both polarizations.  In the ~708-708.5 eV range, in what seems a reversal of the $T<T_M$ case, there is only one distinct peak for $E\parallel$$c$ while now two peaks for $E\perp$$c$.  One of the latter has coincident energy with the $E\parallel$$c$ peak, while the other is a shoulder at lower energy.  For Sn-doped at $T$=300 K, which may be considered a special case since it is near its Morin transition, appears mostly like the $T>T_M$ spectra, but the 708-708.5 eV peaks are not coincident as in the other spectral pairs in this state, and the shoulder of the $E\parallel$$c$ peak at that energy is not visible.  Furthermore there is a very slight shift of the 709.5-710 eV peak which is not present in the others for $T>T_M$.  Thus the Sn-doped $L_{III}$-edge spectra at $T$=300 K seem to mostly, but not fully, correspond to the Morin state spectra exhibited by Zn-doped and undoped samples. These differences suggest inhomogeneity of the magnetic phase near the transition temperature, or that it is not completely a first order phase transition.   \\

Calculations using the multiplet program CTM4XAS \cite{Stavitski10} were performed and are plotted below the experimental curves in Fig.~\ref{fig:L3}.  They were calculated using Fe$^{3+}$ configuration with the $F_{dd}$ Slater integral reduced to 70\% of its Hartree Fock value (or 88\% of its atomic value). The $F_{pd}$ and $G_{dp}$ integrals were used at their atomic value (80\% of Hartree Fock).  Other settings were C3i symmetry, a molecular exchange parameter of 30 meV, and Gaussian broadening of 0.2 eV and Lorentzian broadenings of the first and second main peaks of 0.1 eV and 0.35 eV respectively.  To line up the main peak positions with the experimental spectra, the crystal field parameter of 10$Dq$ was set to 1.2 eV.  The results indicate that the first structure at 708-708.5 eV is dominated by transitions to an empty $t_{2g}$ state (orbitals not facing the oxygen 2$p$ orbital), leading to final state configurations that have the 5 spin-up states plus a $t_{2g}$-down state occupied ($t_{2g}^4e_g^2$). The second structure at 709.5-710 eV is mainly the five spin-up states plus an $e_g$ (facing the oxygen) spin-down state occupied.  Due to the large 2$p$3$d$ interactions, there is a significant mixture of the nature of states, yielding excitations that are distributed among $t_{2g}$, $e_g$, and multi-electron excitations.  Besides the main $t_{2g}$ character of the first peak structure, the composition of its final states also includes 5-15\% $t_{2g}^3e_g^3$ and 0-10\% $t_{2g}^5e_g^1$ configurations, due, respectively, to excitation to an $e_g$ state and configurations that can only be reached via 2-electron excitations.  Likewise, the second peak structure also includes 10-25\% $t_{2g}^4e_g^2$ and 0-10\% $t_{2g}^5e_g^1$ configurations.      While not optimized to reproduce all of the fine features, these simulations do however capture the general trend of the observed polarization dependence of the AF state ($T<T_M$).\\



In summary, the XMLD spectra at the Fe $L_{II}$-edge have clarified the magnetic states of undoped, and 1\% molar Ti, Sn, and Zn substitutions.  These characterizations could provide guidance for the range of temperature and magnetic fields for effective Hall effect measurements \cite{Rettie16}.  $T_M$ of the undoped film, determined by XMLD, was lowered to a value entirely consistent with Ref.~\cite{Simomura15} which used a bulk probe.  This suggests that the magnetic state at the surface and deep into the film are the same.  The donor dopants Ti$^{4+}$ and Sn$^{4+}$ increased $T_M$, in surprising contrast to their effect reported for the bulk.  It would appear as if, in thin films, Ti$^{4+}$ and Sn$^{4+}$ dopants behave as other dopants such as Ir$^{4+}$, whose increase of $T_M$ was attributed to increasing magnetic single ion anisotropy as compared to the competing dipolar term\cite{Liu86}, but the stark difference from the behavior in the bulk is yet unexplained.  Perhaps a greater proportion of electrons donated from Ti in the thin films goes to a more iterant band, rather than being more localized on Fe$^{2+}$ ions and decreasing $T_M$, as is thought for the bulk\cite{Besser67}.  To increase $T_M$, the single ion anisotropy should be increased relative to the dipolar term \cite{Artman65}.  Detailed structure investigation to determine atom positions, in conjunction with microscopic models, would clarify the doping dependence of the anisotropies, as well as guiding parameters for multiplet calculations, which have been shown here to be in good 
agreement with the measured absorption spectra.  The 1\% Zn-doped thin film sample had similar magnetism to that observed in the bulk ceramics\cite{Varshney14}, being in the Morin phase down to at least 5 K.  For the first time, an inversion of the signs of the normalized dichroism peaks was observed by application of a sufficiently strong magnetic field, consistent with previous studies on bulk samples using magnetization and ultrasonic techniques.  This unambiguously confirms the magnetic nature of the measured dichroism.  The variation across samples and temperatures, of features in the Fe $L_{III}$-edge spectra were entirely dependent on the magnetic state.  The room temperature $T_M$ found for the Sn-doped sample may open opportunities for $in$-$situ$ investigations of the effect of magnetic state on the properties of the hematite film, without the need of extreme cooling apparatus or a high magnetic field.  For the PEC application, an abrupt change in the photocurrent either across the transition temperature, or with magnetic field for temperature slightly below it, would clarify the effect (if any) of the spin-ordered state on PEC performance. \\

We gratefully acknowledge helpful discussions with Jaeyoung Kim, Maytal Caspary Toroker, and Amit Keren, and are appreciative of the use of Amit's experimental facilities, and Itzik Kapon for preliminary magnetization tests on our samples.  We thank Helmholtz-Zentrum Berlin (HZB)  for the allocation of synchrotron radiation beamtime at Bessy II Berlin.  D.~S. Ellis thankfully acknowledges the financial support from HZB, and also acknowledges support from The Center for Absorption in Science at the Ministry of Aliyah and Immigrant absorption.  This research has received funding from the European Research Council under the European Union's Seventh Framework programme (FP/200702013) / ERC Grant agreement n. [617516]. D.~A. Grave acknowledges support by Marie-Sklodowska-Curie Individual Fellowship no. 659491.


\end{document}